# A note on coloring line arrangements

Eyal Ackerman[*]    Rom Pinchasi[†]

**Abstract**

We show that the lines of every arrangement of $n$ lines in the plane can be colored with $O(\sqrt{n/\log n})$ colors such that no face of the arrangement is monochromatic. This improves a bound of Bose et al. [1] by a $\Theta(\sqrt{\log n})$ factor. Any further improvement on this bound will improve the best known lower bound on the following problem of Erdős: Estimate the maximum number of points in general position within a set of $n$ points containing no four collinear points.

## 1 Introduction

Let $\mathcal{A}$ be an arrangement of lines in the plane. Denote by $\chi(\mathcal{A})$ the minimum number of colors required for coloring the lines of $\mathcal{A}$ such that there is no monochromatic face in $\mathcal{A}$, that is, the boundary of every face is colored with at least two colors. Bose et al. [1] proved that $\chi(A) \leq O(\sqrt{n})$ for every *simple* arrangement $\mathcal{A}$ of $n$ lines, and that there are arrangements that require $\Omega(\log n/\log\log n)$ colors. We improve their upper bound by a $\Theta(\sqrt{\log n})$ factor, and extend it to not necessarily simple arrangements.

**Theorem 1.** *The lines of every arrangement of $n$ lines in the plane can be colored with $O(\sqrt{n/\log n})$ colors such that no face of the arrangement is monochromatic.*

A set of points in the plane is in *general position* if it does not contain three collinear points. Let $\alpha(S)$ denote the maximum number of points in general position in a set $S$ of points in the plane, and let $\alpha_4(n)$ be the minimum of $\alpha(S)$ taken over all sets $S$ of $n$ points in the plane with no four point on a line. Erdős pointed out that $\alpha_4(n) \leq n/3$ and suggested the problem of determining or estimating $\alpha_4(n)$. Füredi [3] proved that $\Omega(\sqrt{n\log n}) \leq \alpha_4(n) \leq o(n)$.

We observe that any improvement of the bound in Theorem 1 would immediately imply a better lower bound for $\alpha_4(n)$. Indeed, suppose that $\chi(A) \leq k(n)$ for any arrangement of $n$ lines, and let $P$ be a set of $n$ points, no four on a line. Let $P^*$ be the dual arrangement of a slightly perturbed $P$ (according to the usual point-line duality, see, e.g., [2, § 8.2]). Color $P^*$ with $k(n)$ colors such that no face is monochromatic, let $S^* \subseteq P^*$ be the largest color class, and let $S$ be its dual point set. Observe that the size of $S$ is at least $n/k(n)$ and it does not contain three collinear points, since the three lines that correspond to any three collinear points in $P$ bound a face of size three in $P^*$.

## 2 Proof of Theorem 1

Let $\mathcal{A}$ be an arrangement of $n$ lines. We show that $O(\sqrt{n/\log n})$ colors suffice for coloring the lines of $\mathcal{A}$ such that no face in $\mathcal{A}$ is monochromatic. Call a subset of lines *independent* if there is no face that is bounded just by lines from this subset. The proof is based on the following fact.

---

[*]Department of Mathematics, Physics, and Computer Science, University of Haifa at Oranim, Tivon 36006, Israel. ackerman@sci.haifa.ac.il.

[†]Mathematics Department, Technion—Israel Institute of Technology, Haifa 32000, Israel. room@math.technion.ac.il. Supported by BSF grant (grant No. 2008290).

**Theorem 2.** *There is an absolute constant $c > 0$ such that every arrangement of $n$ lines contains at least $c\sqrt{n \log n}$ independent lines.*

We color the lines in $\mathcal{A}$ such that no face is monochromatic by following the same method as in [1]. That is, we iteratively find a large subset of independent lines (whose existence is guaranteed by Theorem 2), color them with the same (new) color, and remove them from $\mathcal{A}$.

Clearly, this algorithm produces a valid coloring. We verify, by induction $n$, that at most $\frac{6}{c}\sqrt{n/\log n}$ colors are used in this coloring. We assume the bound is valid for all $n \leq 256$ (by taking sufficiently small $c > 0$). For $n > 256$, we have $\log 4 < \frac{1}{4}\log n$. Let $i$ be the smallest integer such that after $i$ iterations the number of remaining lines is at most $n/4$. Since in each of these iterations at least $c\sqrt{\frac{n}{4}\log \frac{n}{4}} \geq c\sqrt{\frac{n}{8}\log n}$ vertices are removed, $i \leq \frac{3n/4}{c\sqrt{\frac{n}{8}\log n}} \leq \frac{3}{\sqrt{2}c}\sqrt{n/\log n}$. Therefore, by the induction hypothesis the number of colors that the algorithm uses is at most

$$i + \frac{6}{c}\sqrt{\frac{\frac{n}{4}}{\log \frac{n}{4}}} < \frac{3}{\sqrt{2}c}\sqrt{\frac{n}{\log n}} + \frac{3}{c}\sqrt{\frac{n}{\log n - \frac{1}{4}\log n}} < \frac{3}{\sqrt{2}c}\sqrt{\frac{n}{\log n}} + \frac{2\sqrt{3}}{c}\sqrt{\frac{n}{\log n}} < \frac{6}{c}\sqrt{\frac{n}{\log n}} \quad \square$$

The proof of Theorem 2 is based on a result about independent sets in sparse hypergraphs. A *hypergraph* $H = (V, E)$ consists of a set of vertices $V$ and a set of edges $E \subseteq 2^V$. If the size of every edge is $k$, then $H$ is *k-uniform*. For a set $Z \subseteq V$ the *induced* hypergraph $H[Z]$ consists of $Z$ and the edges of $H$ that are contained in $Z$. A set $I \subseteq V$ is an *independent set* of $H$, if $I$ contains no edge of $H$.

Kostochka et al. [4] proved that every $n$-vertex $(k+1)$-uniform hypergraph in which every $k$ vertices are contained in at most $d < n/(\log n)^{3k^2}$ edges contains an independent set of size $\Omega\left(\left(\frac{n}{d}\log \frac{n}{d}\right)^{1/k}\right)$. In fact, a careful look at their proof reveals the following result, that we stated for 3-uniform hypergraphs, since this is the case that we need.

**Theorem 3** ([4]). *Let $H = (V, E)$ be an $n$-vertex 3-uniform hypergraph, such that every pair of vertices appear in at most $d < n/(\log n)^{12}$ edges. Let $X \subseteq V$ be a random set of vertices chosen independently with probability $p = n^{-2/5}/(d \log \log \log n)^{3/5}$, and let $Z$ be an independent set chosen uniformly at random from the independent sets in $H[X]$. Then with high probability $\mathbf{E}[|Z|] \geq \Omega(\sqrt{n \log n})$.*

With Theorem 3 in hand we can now prove Theorem 2.

*Proof of Theorem 2:* Let $L$ be a set of $n$ lines and let $\mathcal{A}$ be the corresponding arrangement. For a subset $L' \subseteq L$ we say that a face $f$ of $\mathcal{A}$ is *bad* with respect to $L'$ if all the lines bounding $f$ are in $L'$.

Define a 3-uniform hypergraph $H$ whose vertex set are the lines in $\mathcal{A}$ and whose edge set consists of sets of three vertices such that the corresponding lines bound a face of size three in $\mathcal{A}$. Observe that every pair of lines can bound at most four faces of size three, therefore every pair of vertices in $H$ appears in at most four edges. Let $X$ be a random set of vertices (lines) chosen independently with probability $p = n^{-2/5}/(4 \log \log \log n)^{3/5}$. There are $O(n^2)$ faces in $\mathcal{A}$ and $O(n)$ of them are of size two (since every line can bound at most four such faces). Therefore, the expected number of bad faces with respect to $X$ of size different than three is $O(p^2 n + p^4 n^2) = o(\sqrt{n/\log n})$.

It follows from Theorem 3 and Markov's inequality that with high probability the set of lines that correspond to $X$ contains a subset $Z$ of $\Omega(\sqrt{n \log n})$ lines such that there are no faces of size three that are bad with respect to $Z$, and the number of bad faces of size different than three is $o(\sqrt{n/\log n})$. By removing from $Z$ a vertex from each bad face we obtain an independent set of lines of size $\Omega(\sqrt{n \log n})$. $\square$

# References

[1] P. Bose, J. Cardinal, S. Collette, F. Hurtado, S. Langerman, M. Korman, and P. Taslakian, Coloring and guarding arrangements, *28th European Workshop on Computational Geometry (EuroCG)*, March 19–21, 2012, Assisi, Perugia, Italy. Also available at: `http://arxiv.org/abs/1205.5162`

[2] M. de Berg, O. Cheong, M. van Krefeld, M. Overmars, *Computational Geometry: Algorithms and Applications*, 3rd edition, Springer, 2008.

[3] Z. Füredi, Maximal independent subsets in Steiner systems and in planar sets, *SIAM J. Disc. Math.* **4** (1991), 196–199.

[4] A. Kostochka, D. Mubayi, J. Versatraete, On independent sets in hypergraphs, *Random Structures and Algorithms*, to appear.